\documentclass[twocolumn]{aastex63}

\usepackage{rotating}
\usepackage{acronym}
\usepackage{xspace}
\usepackage{multirow}
\usepackage{mathtools}
\usepackage{amsmath}
\usepackage{hyperref}
\usepackage{units}

\usepackage{xcolor}

\graphicspath{{./}{figures/}}

\definecolor{chmagenta}{rgb}{0.54, 0.17, 0.88}

\def\mchirp{\ensuremath{\mathcal{M}_\mathrm{c}}\xspace}
\def\Mtot{\ensuremath{M_\mathrm{tot}}\xspace}

\def\q{\ensuremath{q}\xspace}
\def\z{\ensuremath{z}\xspace}

\def\pdet{\ensuremath{p_\mathrm{det}}\xspace}

\def\Msun{\ensuremath{\mathit{M_\odot}}\xspace}

\def\kms{\ensuremath{\mathrm{km\,s^{-1}}}\xspace}

\def\OneG{\ensuremath{\texttt{1G}}\xspace}
\def\TwoG{\ensuremath{\texttt{2G}}\xspace}
\def\OneGOneG{\texttt{1G+1G}\xspace}
\def\NGOneG{\texttt{NG+1G}\xspace}
\def\NGNG{\texttt{NG+NG}\xspace}
\def\NGMG{\texttt{NG+$\leq$NG}\xspace}

\def\LVC{\texttt{LVC}\xspace}
\def\IMF{\texttt{IMF}\xspace}

\def\Nmerge{\ensuremath{N_\mathrm{merge}}\xspace}
\def\vesc{\ensuremath{V_\mathrm{esc}}\xspace}
\def\NOneG{\ensuremath{N_\mathrm{1G}}\xspace}
\def\Nbudget{\ensuremath{N_\mathrm{budget}}\xspace}
\def\Mf{\ensuremath{M_\mathrm{f}}\xspace}
\def\Chif{\ensuremath{\chi_\mathrm{f}}\xspace}
\def\vk{\ensuremath{V_\mathrm{k}}\xspace}

\def\MedMassThresh{\ensuremath{100\,M_\odot}\xspace}
\def\MedMassAboveThreshNGOneGLVC{\ensuremath{\mathrm{nine}}\xspace}
\def\MedMassAboveThreshNGMGLVC{\ensuremath{\mathrm{six}}\xspace}
\def\MedMassAboveThreshNGNGLVC{\ensuremath{\mathrm{four}}\xspace}

\def\MedSpinThresh{\ensuremath{0.5}\xspace}
\def\MedSpinBelowThreshNGOneGLVC{\ensuremath{8}\xspace}
\def\MedSpinBelowThreshNGMGLVC{\ensuremath{14}\xspace}

\def\ZeroSpinKickPeak{\ensuremath{0.36}\xspace}
\def\MedKickThresh{\ensuremath{300}\xspace}
\def\MedKickBelowThresh{\ensuremath{\sim 3\mbox{--}4}\xspace}
\def\KickRangeNGNG{\ensuremath{\sim 100\mbox{--}2100}\xspace}

\def\fretOneThousandFiveMergersNGNGLVC{\ensuremath{18\%}\xspace}
\def\fretFiveHundredFiveMergersNGNGLVC{\ensuremath{1.2\%}\xspace}
\def\fretThreeHundredFiveMergersNGNGLVC{\ensuremath{0.06\%}\xspace}
\def\fretOneThousandFiveMergersNGNGIMF{\ensuremath{19\%}\xspace}
\def\fretFiveHundredFiveMergersNGNGIMF{\ensuremath{1.7\%}\xspace}
\def\fretThreeHundredFiveMergersNGNGIMF{\ensuremath{0.13\%}\xspace}

\def\fretMinOneThousandNGOneGLVC{\ensuremath{48\%}\xspace}
\def\fretMinFiveHundredNGOneGLVC{\ensuremath{8\%}\xspace}
\def\fretMinThreeHundredNGOneGLVC{\ensuremath{1\%}\xspace}
\def\fretMinOneThousandNGOneGIMF{\ensuremath{59\%}\xspace}
\def\fretMinFiveHundredNGOneGIMF{\ensuremath{13\%}\xspace}
\def\fretMinThreeHundredNGOneGIMF{\ensuremath{2\%}\xspace}

\def\NGOneGGreaterEightyThreeHundred{\ensuremath{34\%}\xspace}
\def\NGOneGGreaterEightyFiveHundred{\ensuremath{61\%}\xspace}
\def\NGOneGGreaterEightyOneThousand{\ensuremath{78\%}\xspace}

\def\NGOneGGreaterEightyThreeHundredDet{\ensuremath{52\%}\xspace}
\def\NGOneGGreaterEightyFiveHundredDet{\ensuremath{76\%}\xspace}
\def\NGOneGGreaterEightyOneThousandDet{\ensuremath{87\%}\xspace}

\def\LVCLessThanOneHundred{\ensuremath{<\,1.5\%}\xspace}

\def\FracAboveOneHundredMsunNGNGIMFThreeHundred{\ensuremath{1.6\%}\xspace}

\def\FracAboveOneHundredMsunNGNGLVCFiveHundred{\ensuremath{9.6\%}\xspace}
\def\FracAboveOneHundredMsunNGMGLVCFiveHundred{\ensuremath{29\%}\xspace}
\def\FracAboveOneHundredMsunNGOneGLVCFiveHundred{\ensuremath{55\%}\xspace}

\def\FracAboveOneHundredMsunNGMGLVCOneThousandBHBudgetOneHundred{\ensuremath{61\%}\xspace}
\def\FracAboveOneHundredMsunNGMGLVCOneThousandBHBudgetThirty{\ensuremath{48\%}\xspace}
\def\FracAboveOneHundredMsunNGMGLVCOneThousandBHBudgetTen{\ensuremath{11\%}\xspace}

\def\MassesAboveOneHundredMsunFiveHundredRange{\ensuremath{7\%\mbox{--}55\%}\xspace}
\def\FracAboveOneHundredMsunNGOneGOneThousandIMF{\ensuremath{69\%}\xspace}
\def\AGNFirstGenerationToHierarchical{\ensuremath{46}\xspace}
\acrodef{GW}{gravitational-wave}
\acrodef{BH}{black hole}
\acrodef{BBH}{binary black hole}
\acrodef{NS}{black hole}
\acrodef{BNS}{binary neutron star}
\acrodef{IMBH}{intermediate-mass black hole}

\acrodef{LIGO}{Laser Interferometer Gravitational-wave Observatory}
\acrodef{LVC}{LIGO Scientific and Virgo Collaboration}
\acrodef{O1}{first observing run}
\acrodef{O2}{second observing run}
\acrodef{O3}{third observing run}
\acrodef{O3a}{first half of the third observing run}

\acrodef{SNR}{signal-to-noise ratio}
\acrodef{PSD}{power spectral density}
\acrodef{NR}{numerical relativity}

\acrodef{GC}{globular cluster}
\acrodef{NC}{nuclear cluster}
\acrodef{AGN}{active galactic nucleus}

\acrodef{PI}{pair-instability}
\acrodef{IMF}{initial mass function}

\acrodef{CDF}{cumulative distribution function}

\newcommand{\KICP}{\affiliation{Kavli Institute for Cosmological Physics, The University of Chicago, 5640 South Ellis Avenue, Chicago, Illinois 60637, USA}}
\newcommand{\EFI}{\affiliation{Enrico Fermi Institute, The University of Chicago, 933 East 56th Street, Chicago, Illinois 60637, USA}}
\newcommand{\UChicago}{\affiliation{Department of Physics, Department of Astronomy \& Astrophysics, The University of Chicago, 5640 South Ellis Avenue, Chicago, Illinois 60637, USA}}

\shorttitle{Avoiding a Cluster Catastrophe}
\shortauthors{Zevin \& Holz 2022}

\begin{document}

\title{Avoiding a Cluster Catastrophe: \\Retention Efficiency and the Binary Black Hole Mass Spectrum
}

\author[0000-0002-0147-0835]{Michael~Zevin}\email{michaelzevin@uchicago.edu}\thanks{NASA Hubble Fellow}
\KICP \EFI
\author[0000-0002-0175-5064]{Daniel~E.~Holz}
\KICP\EFI\UChicago

\begin{abstract}
The population of binary black hole mergers identified through gravitational waves has uncovered unexpected features in the intrinsic properties of black holes in the universe. 
One particularly surprising and exciting result is the possible existence of black holes in the pair-instability mass gap, $\sim50\mbox{--}120~M_\odot$. Dense stellar environments can populate this region of mass space through hierarchical mergers, with the retention efficiency of black hole merger products strongly dependent on the escape velocity of the host environment. 
We use simple toy models to represent hierarchical merger scenarios in various dynamical environments. 
We find that hierarchical mergers in environments with high escape velocities ($\gtrsim300~\kms$) are efficiently retained. 
If such environments dominate the binary black hole merger rate, this would lead to an abundance of high-mass mergers that is potentially incompatible with the empirical mass spectrum from the current catalog of binary black hole mergers. 
Models that efficiently generate hierarchical mergers, and contribute significantly to the observed population, must therefore be tuned to avoid a ``cluster catastrophe'' of overproducing binary black hole mergers within and above the pair-instability mass gap. 
\end{abstract}

%\keywords{space!}

\section{Introduction}\label{sec:intro}

The population of \ac{BBH} mergers identified by the LIGO--Virgo \ac{GW} interferometer network has uncovered interesting and unexpected features in the intrinsic properties of \acp{BH} in the Universe. 
The \ac{LVC}~\citep{GW190521,GWTC2,GWTC3}, as well as groups external to the \ac{LVC}~\citep{Venumadhav2020,Nitz2021a}, has reported a number of component \acp{BH} with significant posterior support for masses $\gtrsim50~\Msun$, in tension with predictions for the mass limit imposed by the \ac{PI} process~\citep{Woosley2017,Woosley2019,Marchant2019,Renzo2020}. 
Though uncertainties in the underlying physical processes that lead to this phenomenon~\citep{Farmer2020,Woosley2021} and alternative astrophysical prior interpretations~\citep{Fishbach2020a,Nitz2020a} put into question the putative observations of \acp{BH} in the \ac{PI} mass gap, mergers in dense stellar environments offer a natural means of polluting this anticipated dearth in the \ac{BH} mass spectrum. 

In dense stellar environments such as \acp{GC}, \acp{NC}, and \ac{AGN} disks, \ac{BH} progenitors and \acp{BH} themselves dynamically interact and merge with other members of the cluster. 
Merging stars may maintain low enough He-core masses to avoid disruption through \ac{PI}, leading to \acp{BH} that have masses well above the \ac{PI} limit theorized for massive stars~\citep{Spera2019,DiCarlo2019,DiCarlo2021,Kremer2020a,Banerjee2021a,Tagawa2021c,Gonzalez2021}. 
Alternatively, \acp{BH} that merge within the cluster can be retained in the cluster environment and proceed to merge again with other members of the cluster, leading to increasingly massive \ac{BH} mergers. 
This merger pathway, known as hierarchical mergers, has been shown to efficiently pollute the \ac{PI} mass gap~\citep{Fishbach2017,Gerosa2017a,Gerosa2019a,Rodriguez2019,Yang2019a,Kimball2020,Doctor2021,Fragione2022c,Mapelli2021,Mapelli2022}, and the occurrence of hierarchical mergers across various dynamical environment may help to explain the origin of certain \ac{GW} events~\citep[e.g.,][]{Rodriguez2020,Baibhav2021,Fragione2021,Gayathri2021,Gerosa2021b,Kimball2021,Tagawa2021,Wong2021a,Zevin2021}. 
This process has also been invoked to explain the growth of \acp{IMBH} in cluster environments~\citep{Quinlan1987} and might be assisted by repeated stellar mergers prior to the collapse of the \ac{IMBH} seed~\citep[e.g.,][]{PortegiesZwart2002,Gurkan2004,Giersz2015,Freitag2006,Kremer2020a,DiCarlo2021,Gonzalez2021,Fragione2022,Fragione2022b}. 

The efficiency of this channel is highly dependent on the global properties of a cluster environment; when two \acp{BH} merge, anisotropies in \ac{GW} emission impart momentum to the merger product known as a gravitational recoil kick or radiation rocket~\citep{Favata2004}, which is amplified by asymmetries in the masses and spins of the system. 
If this kick exceeds the escape velocity of their host environment, the \ac{BH} merger product will be ejected and will be incapable of encountering another \ac{BH} with which to merge again~\citep{Merritt2004}. 
Higher-density environments such as \acp{NC} and \acp{AGN} will thus be much more efficient at retaining merger products than lower-mass \acp{GC}~\citep{Gerosa2019a}, and the $\mathcal{O}(10~\kms)$ escape velocities of young star clusters and open clusters will hardly be able to retain any merger products whatsoever. 
Inference on the recoil kick velocities of \ac{BBH} mergers has found that an appreciable number of merger products from recent \ac{GW} catalogs should be retained by their host environments if they merged in star clusters~\citep{Mahapatra2021}. 

A natural inclination would be to attribute the higher-mass systems observed by the \ac{LVC} to such high-mass and high-density environments. 
However, this interpretation must be exercised with caution. 
If a particular environment is too efficient at retaining merger products, it may lead to runaway \ac{BH} mergers and a cluster catastrophe~\citep{Fishbach2017}. 
In this case, the predicted mass spectrum may have an overabundance of high-mass mergers that is inconsistent with the observed masses of \ac{GW} events. 
Ground-based \ac{GW} interferometers are more sensitive to higher-mass \ac{BH} mergers (up to a certain point), and selection effects must be accounted for when comparing \ac{BH} population predictions to the catalog of observed systems. 

In this paper, we use simple models for hierarchical mergers to demonstrate how the expected mass spectrum of hierarchical mergers is impacted by aspects of the host environment, with a particular focus on how environments with high retention efficiencies can potentially produce an overabundance of high-mass mergers that are inconsistent with the empirical \ac{BH} mass spectrum. 
The relative lack of high-mass merger observations can thereby place constraints on the contribution of hierarchical mergers from certain dynamical environments to the full \ac{BBH} population. 
In Section~\ref{sec:methods} we outline our models and model assumptions, and discuss how hierarchical merger trees are constructed. 
We present the results from our models in Section~\ref{sec:results}, highlighting the impact of escape velocity on the predicted \ac{BH} mass spectrum. 
We conclude with implications and caveats of our analysis in Section~\ref{sec:discussion}. 

\section{Seeding, Growing, and Pruning Hierarchical Merger Trees}\label{sec:methods}

The primary goal of this work is to investigate how host cluster properties, in particular escape velocities, impact the observed mass spectrum of hierarchically merging \acp{BBH}. 
We pre-generate merger trees that start with first-generation (\OneG) \ac{BH} seeds and grow them by merging \acp{BH} in series while tracking pertinent properties of their remnants. 
A number of different pairing assumptions for hierarchical mergers are used as a representation of the pairing processes in dynamical formation environments of \acp{BBH}. 
We then prune the merger trees to account for the escape velocity of host environments as well as the budget of \acp{BH} available in a particular environment. 
Finally, we apply selection effects to a compiled population of hierarchical mergers and compare the resulting mass distributions to those observed by LIGO--Virgo. 
Our models are similar in construction to those in \cite{Gerosa2021a}. 
The codebase for performing our analysis is available on Github\footnote{\url{https://github.com/michaelzevin/hierarchical-mergers}}, with data products used in our analyses available on Zenodo.\footnote{\cite{cluster_catastrophe_dataset}, \url{https://zenodo.org/record/6811921}}

\subsection{Initial Population of Black Holes}\label{subsec:first_gen}
To seed the hierarchical merger trees, we require an initial population of \acp{BH} parameterized by their component masses, component spin magnitudes, and birth redshifts. 
In our default model, we draw component \ac{BH} masses for the seed \ac{BBH} merger using the joint mass and mass ratio posterior predictive distribution from the \texttt{Power Law + Peak} model of \cite{GWTC3_pops}. 
Spin magnitudes are drawn from the \texttt{Default} model of \cite{GWTC3_pops}. 
We refer to this model as \LVC. 

Though the mass distribution measured by the \ac{LVC} provides our best empirical constraints for the merging \ac{BBH} mass distribution, this measured distribution may not be representative of the \OneG population in the dynamical environments we consider. 
This is because the observed population of mergers may itself include hierarchical mergers, and thus the fits by the \ac{LVC} do not represent the \OneG population explicitly; any model in which higher-generational mergers account for some portion of the observed \acp{BBH} mergers may have an excess of massive mergers by construction. 
For this reason, similar to \cite{Gerosa2021a}, we also consider a \OneG mass distribution that follows the \ac{IMF} of massive stars, $p(m) \propto m^{-2.3}$~\citep{Kroupa2001}, and draw both \OneG component \acp{BH} according to this distribution between the bounds $m \in [5 \Msun, 50 \Msun]$. 
Assuming efficient angular momentum transport and low \ac{BH} birth spins for the \OneG population~\citep[e.g.,][]{Spruit2002,Fuller2019a}, we assign \ac{BH} spin magnitudes uniformly between $a \in [0,0.1]$. 
This \OneG model is referred to as \IMF. 

Because all the \acp{BBH} we consider are dynamically formed, we assume that spin orientations of both components relative to the orbital angular momentum are distributed isotropically on the sphere. 
Redshifts of the first merger are drawn uniformly in comoving volume between $z \in [0,5]$. 

\subsection{Synthesizing Merger Trees}\label{subsec:merger_trees}

Branches of a merger tree represent a single chain of hierarchical mergers. 
Each branch is initially grown with no limitations from escape velocity or an assumed maximum number of \acp{BH} that can partake in the hierarchical assembly. 
We assume three distinct pairing methods for growing each individual hierarchical merger branch: 
\begin{enumerate}
\item \NGOneG: The merger product being tracked continually merges with other \acp{BH} from the \OneG distribution. 
As hierarchical mergers produce successive populations of larger \acp{BH}, we repeatedly and exclusively pair these NG \acp{BH} with either the primary or secondary \ac{BH} from the original (\OneG) distribution. 
This method resembles the expectations of a runaway merger scenario, which has been invoked for forming \acp{IMBH}~\citep{Quinlan1987,Miller2002}, as well as the assembly of hierarchical \acp{BH} in the migration traps of \ac{AGN} disks~\citep[e.g.,][]{Mckernan2014,Yang2019a,McKernan2020,Tagawa2021,Li2022}. 
\item \NGNG: The merger product being tracked continually merges with another \ac{BH} that has gone through the same number of prior mergers. 
This expedites the buildup in mass of the merger product and resembles the expectations from dense stellar environments where the most massive objects mass-segregate and dominate the dynamics of the dense cluster core, preferentially kicking out lighter components during strong few-body gravitational encounters. 
\item \NGMG: The merger product being tracked merges with a \ac{BH} that has gone through a number of mergers $M \leq N$. 
The probability that a given merger generation $M$ is chosen for the companion is proportional to the number of \OneG \acp{BH} required to synthesize it, $p(M) \propto 2^{-(M-1)}$ (for example, the NG \ac{BH} is twice as likely to merge with a \OneG \ac{BH} than a \TwoG \ac{BH}, and four times as likely to merge with a \OneG \ac{BH} than a 3G \ac{BH}). 
If the merger partner is chosen to be a \OneG \ac{BH}, in the \LVC model we once again split up the primary and secondary components and randomly choose one. 
This is representative of a steady-state limit and is a useful model for means of comparison with the other pairing models.  
\end{enumerate}
Note that the term ``generation'' becomes ambiguous once one exceeds \TwoG mergers; a 3G merger in the \NGOneG channel requires four \OneG \acp{BH}, whereas a 3G merger in the \NGNG channel requires $2 \times 2^3 = 16$ \OneG \acp{BH}. 
Thus, for the majority of this work, we will instead refer to the number of mergers, \Nmerge, that have occurred at a particular point along a branch. 

Our simple model also requires a prescription to determine the amount of time that passes between subsequent mergers. 
This will impact the number of hierarchical mergers that can occur before the present day and potentially be detected via \acp{GW}. 
For our fiducial model, we assume that delay times between each subsequent merger, ${\Delta t = t_{N_\mathrm{merge}+1} - t_{N_\mathrm{merge}}}$, are drawn from a log-uniform distribution between 10 and 100 Myr. 
Merger redshifts are determined assuming Planck cosmological parameters~\citep{PlanckCollaboration2018}. 

Our choice of delay-time distribution is representative of young massive clusters and \acp{GC}, in which \acp{BH} have relatively short mass segregation timescales of $\sim10\mbox{--}100~\mathrm{Myr}$~\citep[e.g.,][]{PortegiesZwart2010}. 
We note that this is a simplifying assumption because the dynamical friction timescale can vary vastly between different host environments~\citep{Antonini2019a,Fragione2020,Fragione2022c}, and other functional forms for the delay times between subsequent mergers may lead to differences in our underlying and detectable populations. 
In Appendix \ref{app:delay_times}, we test the sensitivity of our main results to variations in the assumed delay-time distribution, using extended log-uniform distributions between $[10~\mathrm{Myr},\ 1~\mathrm{Gyr}]$ and $[100~\mathrm{Myr},\ 14~\mathrm{Gyr}]$. 
Though the first of these variations has little impact on results, the second variation acts to suppress the highest mass mergers in our hierarchical scenarios because these will more readily occur at times beyond the present day; see Figure~\ref{fig:escape_velocity_mass_cdf_Tdel_LVC} in Appendix~\ref{app:delay_times}. 

For each merger, we determine the mass of the merger product \Mf, the spin of the merger product \Chif, and the \ac{GW} recoil kick velocity \vk using the \texttt{precession} package~\citep{Gerosa2016}. 
We also track the number of \OneG \acp{BH} needed to generate each merger product, defined as \NOneG (e.g., for the merger product of a \TwoG+\TwoG merger, $\NOneG=4$ since $4$ \OneG \acp{BH} are utilized). 
We grow $5 \times 10^5$ such branches for each merger tree, allowing for 10, 20, and 30 subsequent mergers on each branch for the \NGNG, \NGMG, and \NGOneG scenarios, respectively,\footnote{Merger products above these chosen maximum number of mergers will typically be too massive to be detected by current ground-based \ac{GW} detectors, but could prove important for future \ac{GW} detectors that probe lower frequency regimes and could help constrain rates of \ac{IMBH} mergers formed through hierarchical merger scenarios~\citep[e.g.,][]{Fragione2022c}.} and combine their results when synthesizing our populations. 
The evolution of relevant parameters as a function of \Nmerge is shown in Figure~\ref{fig:merger_product_evolution}.

\subsection{Pruning Merger Trees}\label{subsec:prune_trees}

Once the full merger trees are grown, we proceed to prune the trees using cuts based on both escape velocity, \vesc, and/or \ac{BH} budget, \Nbudget. 
At any point along a given branch, if $\vk \geq \vesc$ or $\NOneG > \Nbudget$, we drop all subsequent mergers that occur since the merger product was either ejected from the environment, or the merger product required more \OneG \acp{BH} than are available. 
In addition, we remove any mergers that occur in the future, i.e. with lookback times $t_\mathrm{lb} < 0$. 
These postprocessing steps account for the environment in which the hierarchical mergers are occurring, which is parameterized by only \vesc and \Nbudget for simplicity. 
Throughout the majority of this work, we leave the budget of \acp{BH} available for hierarchical assembly unconstrained and focus primarily on differing values for \vesc. 
Different pairing scenarios lead to a different total number of \OneG \acp{BH} utilized in the hierarchical assembly process, with the most potential \OneG \acp{BH} used in the \NGNG pairing ($2^{10}$, because we allow for a maximum of 10 subsequent mergers in this scenario). 
The impact of different assumptions for \Nbudget on the resultant hierarchical \ac{BH} mass spectrum is explored in Appendix~\ref{app:BH_budget}.

\subsection{Selection Effects}\label{subsec:selection_effects}

Finally, we incorporate selection effects on the remaining population as a means to compare our models with the empirical distribution of \acp{BBH} mergers. 
Our semianalytic treatment of selection effects uses a precomputed grid of LIGO--Virgo detection probabilities, \pdet,
over chirp mass $\mchirp = (m_1 m_2)^{3/5} / (m_1+m_2)^{1/5}$, mass ratio $\q = m_2/m_1$ with $m_2 \leq m_1$, and redshift \z where $m_1$ and $m_2$ are the primary and secondary masses, respectively~\citep{Zenodo_selection_effects}. 
We ignore spin effects and possible eccentricity in the detectability calculations. 
Our grids assume a three-detector network consisting of LIGO--Livingston, LIGO--Hanford, and Virgo operating at \texttt{midhigh/latelow} sensitivity~\citep{LVC_ObservingScenarios}, with a network detection threshold of ${\mathrm{signal\mbox{-}to\mbox{-}noise\ ratio} > 10}$. 
Relative detection weights for each merger also account for surveyed space-time volume: 
\begin{equation}
    w^i \propto \frac{\pdet^i}{1+z^i} \frac{dV_c}{dz}(z^i)
\end{equation}
where $dV_c/dz$ is the differential comoving volume at redshift $z^i$ and $(1+z)^{-1}$ accounts for time dilation between the merger and the detectors. 
We assume a primary mass detection cutoff of $m_{1,\mathrm{det}}^\mathrm{max} = 500 \Msun$---any system with a (detector-frame) primary mass $m_1 > m_{1,\mathrm{max}}$ has a detection probability of zero, because beyond this the detector spectral sensitivity drops significantly~\citep[e.g.,][]{Mehta2022}.

%%% FIGURE 1
\begin{figure*}[t]
\includegraphics[width=1.0\textwidth]{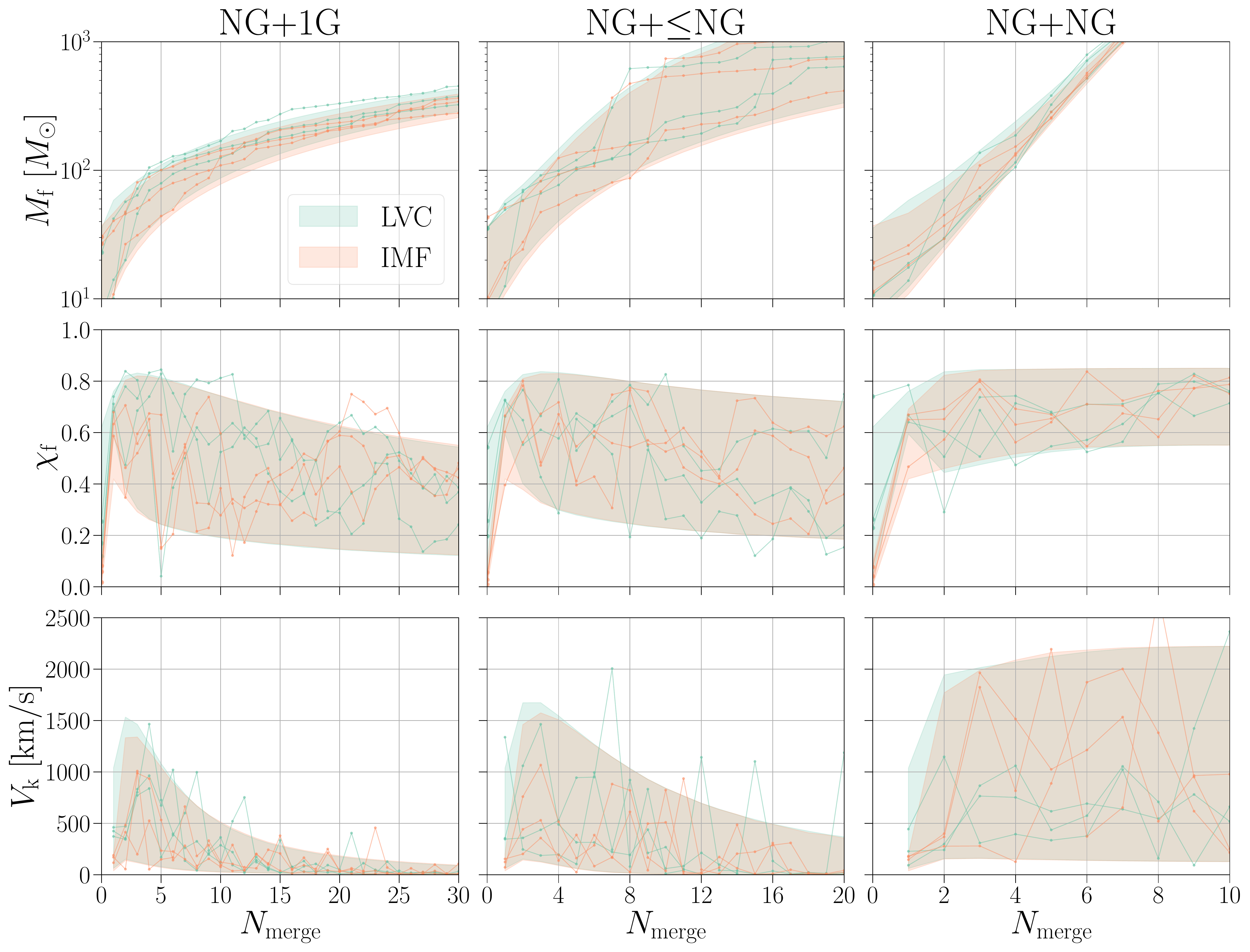}
\caption{Evolution of merger product mass ($\Mf$, top row), dimensionless spin magnitude ($\Chif$, middle row), and recoil kick velocity ($V_\mathrm{k}$, bottom row) of merger products that have proceeded through \Nmerge mergers. 
The columns show the evolution of the three different pairing channels described in Section \ref{subsec:merger_trees}, with colors showing either the \LVC (green) or \IMF (orange) \OneG populations. 
Colored bands contain 90\% of systems, with lines showing representative randomly selected merger histories. 
At $\Nmerge = 0$ we plot the mass and spin values for the primary component of the initial binary and do not show a value for the recoil kick velocity because the binary has yet to merge. 
Merger product masses grow exponentially in the \NGNG pairing scenario and the distribution of recoil kicks remains broad, whereas the mass growth flattens out in the intergenerational \NGOneG and \NGMG pairing scenarios, with the recoil kick distribution approaching $0~\kms$ as a function of \Nmerge because the mass ratio of subsequent mergers enters the regime of $q \ll 1$ in these pairing scenarios. 
Note that the range of \Nmerge plotted on the horizontal axis varies between pairing methods. 
}
\label{fig:merger_product_evolution}
\end{figure*}

%%% FIGURE 2
\begin{figure*}[t]
\includegraphics[width=1.0\textwidth]{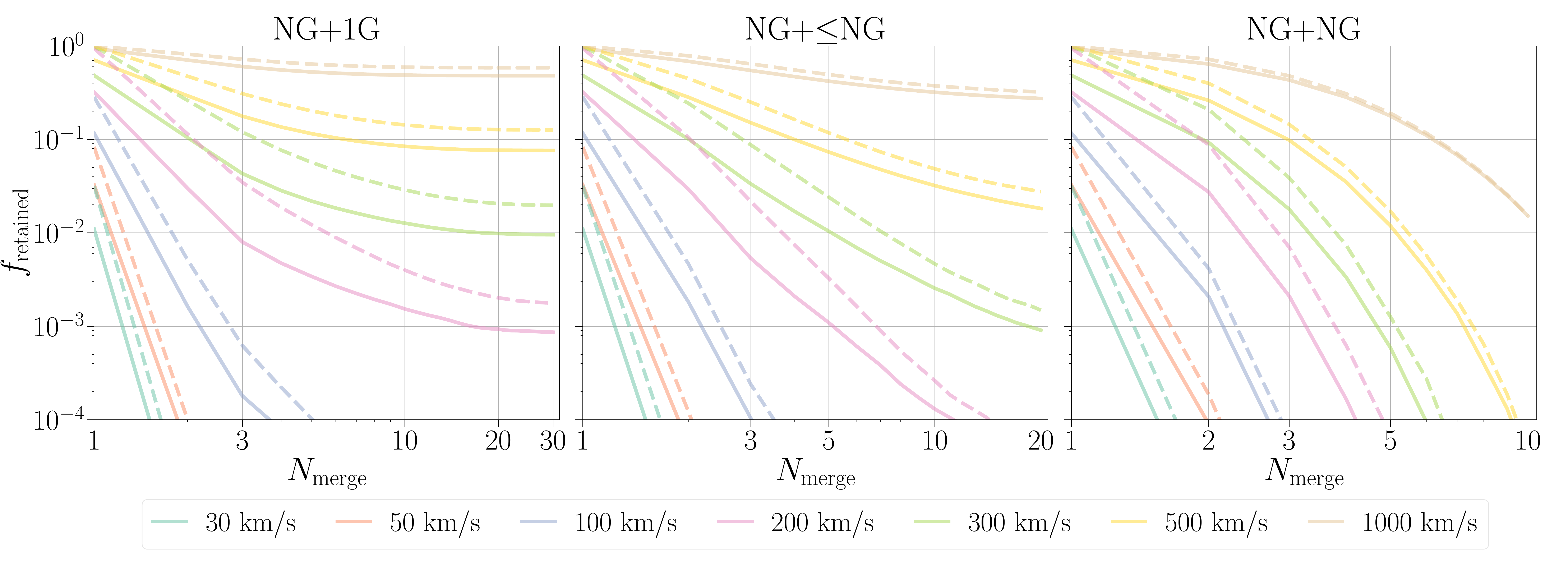}
\caption{Fraction of merger products retained as a function of the number of mergers in the hierarchical assembly process. 
Colored lines indicate various assumptions for escape velocities, and solid and dashed lines indicate the \LVC and \IMF \OneG population, respectively. 
Retention fractions always decrease as a function of \Nmerge in the \NGNG pairing scenario, whereas for escape velocities $\geq 200~\kms$ the retention fractions in the \NGOneG pairing scenario asymptote to specific values as a function of \Nmerge; once hierarchical merger products enter this regime, recoil kicks never eject them from their environment. 
Note that the range of \Nmerge plotted on the horizontal axis varies between pairing methods. 
}
\label{fig:retention_fraction}
\end{figure*}

\section{Results}\label{sec:results}

We first examine general features of the hierarchical merger trees without incorporating constraints based on the host environment: We assume no merger products are ejected from their host environment and that there is an infinite budget of \acp{BH} with which to merge (see Figure~\ref{fig:merger_product_evolution}). 
Merger product masses grow exponentially for the \NGNG pairing, approximately doubling in mass with each subsequent merger, whereas the relative mass growth flattens in both the \NGMG and \NGOneG cases. 
The merger trees that result from the \IMF initial mass distribution push to slightly lower masses than the \LVC initial mass distribution because it has no support for component masses above $50~\Msun$ in the \OneG population. 
The median source-frame mass of the merger product exceeds \MedMassThresh after \MedMassAboveThreshNGOneGLVC, \MedMassAboveThreshNGMGLVC, and \MedMassAboveThreshNGNGLVC mergers for both the \LVC and \IMF initial mass distributions in the \NGOneG, \NGMG, and \NGNG cases, respectively. 

Merger product spins rise to $\sim 0.7$ after the first merger due to the orbital angular momentum of the merging binary~\citep{Fishbach2017}, with a dispersion due to asymmetries in the component masses and the spin orientations of the mergers. 
The distribution of merger product spins remains peaked at $\sim 0.7$ for the \NGNG pairing scenario because mergers are typically near equal mass. 
However, average merger product spins decrease as a function of \Nmerge in the \NGOneG and \NGMG cases; though aligned and antialigned spins are equally likely in an isotropic spin distribution, counterrotating orbits are more efficient at extracting angular momentum than corotating orbits are at depositing it~\citep{Hughes2003}. 
To first order, the decrease in \ac{BH} spin is proportional to the mass ratio of the binary and thus on average the spin distribution of merger products will approach zero after many mergers with asymmetric masses~\citep[e.g.,][]{Fragione2022c,Gerosa2021a}. 
For \NGOneG and \NGMG, the spin of the merger products drops below \MedSpinThresh after $\sim\MedSpinBelowThreshNGOneGLVC$ and $\sim\MedSpinBelowThreshNGMGLVC$ mergers, respectively. 
Though this work focuses on mass distributions through hierarchical assembly, a population of hierarchical mergers will also display distinctive spin signatures, such as high spins that are oriented in the hemisphere opposite of the orbital angular momentum, which may also be used to constrain the abundance of hierarchical \ac{BBH} mergers in the general population~\citep[see, e.g.,][]{Fishbach2022}. 

Recoil kick velocities, as well as the dispersion of the recoil kick velocity distribution, also decrease as a function of \Nmerge in the \NGOneG and \NGMG pairings. 
Though mergers with nonspinning components will have a maximum kick at a mass ratio of \ZeroSpinKickPeak~\citep[e.g.][]{Favata2004}, the strength of the recoil kick drops precipitously as the mass ratio approaches zero even if the component \acp{BH} have significant spin. 
For \NGOneG and \NGMG pairings, median recoil kicks drop below $\MedKickThresh~\kms$ after \MedKickBelowThresh mergers. 
However, recoil kicks stay relatively strong with a broad distribution ranging from $\KickRangeNGNG~\kms$ in the \NGNG case due to its preference for near-equal-mass pairings.

\subsection{Imposing Escape Velocities}\label{subsec:results_escape_velocities}

Escape velocities in dynamical environments that potentially harbor \acp{BH} can range from $\mathcal{O}(10~\kms)$ for present-day \acp{GC} to $\mathcal{O}(100~\kms)$ for \acp{NC}, and up to ${\sim 1000~\kms}$ in \ac{AGN} disks. 
Though host escape velocities can evolve significantly in time (for example, present-day \acp{GC} may have been a few times more massive at birth with escape velocities of several hundred $\kms$, \citealt{Webb2015}), we postprocess our merger trees assuming fixed escape velocities that are representative of different dynamical environments at specific points in their evolution. 
The fraction of retained merger products for various assumed escape velocities is shown in Figure~\ref{fig:retention_fraction}. 

For the \NGNG pairing scenario, the retention fraction drops precipitously as the number of subsequent mergers increases due to near-equal-mass mergers continuously receiving large kicks (see Figure~\ref{fig:merger_product_evolution}). 
Environments with exceptionally large escape velocities can still retain a number of subsequent mergers in this scenario. 
For escape velocities $\gtrsim 1000~\kms$ and the \LVC (\IMF) \OneG populations, $\fretOneThousandFiveMergersNGNGLVC$ ($\fretOneThousandFiveMergersNGNGIMF$) of hierarchical \ac{BBH} systems can be retained in their host environment even after they have gone through five prior mergers, whereas this number drops to $\fretFiveHundredFiveMergersNGNGLVC$ ($\fretFiveHundredFiveMergersNGNGIMF$) for escape velocities of $500~\kms$ and $\fretThreeHundredFiveMergersNGNGLVC$ ($\fretThreeHundredFiveMergersNGNGIMF$) for escape velocities of $300~\kms$.

%%% FIGURE 3

\begin{figure*}[t]
\includegraphics[width=1.0\textwidth]{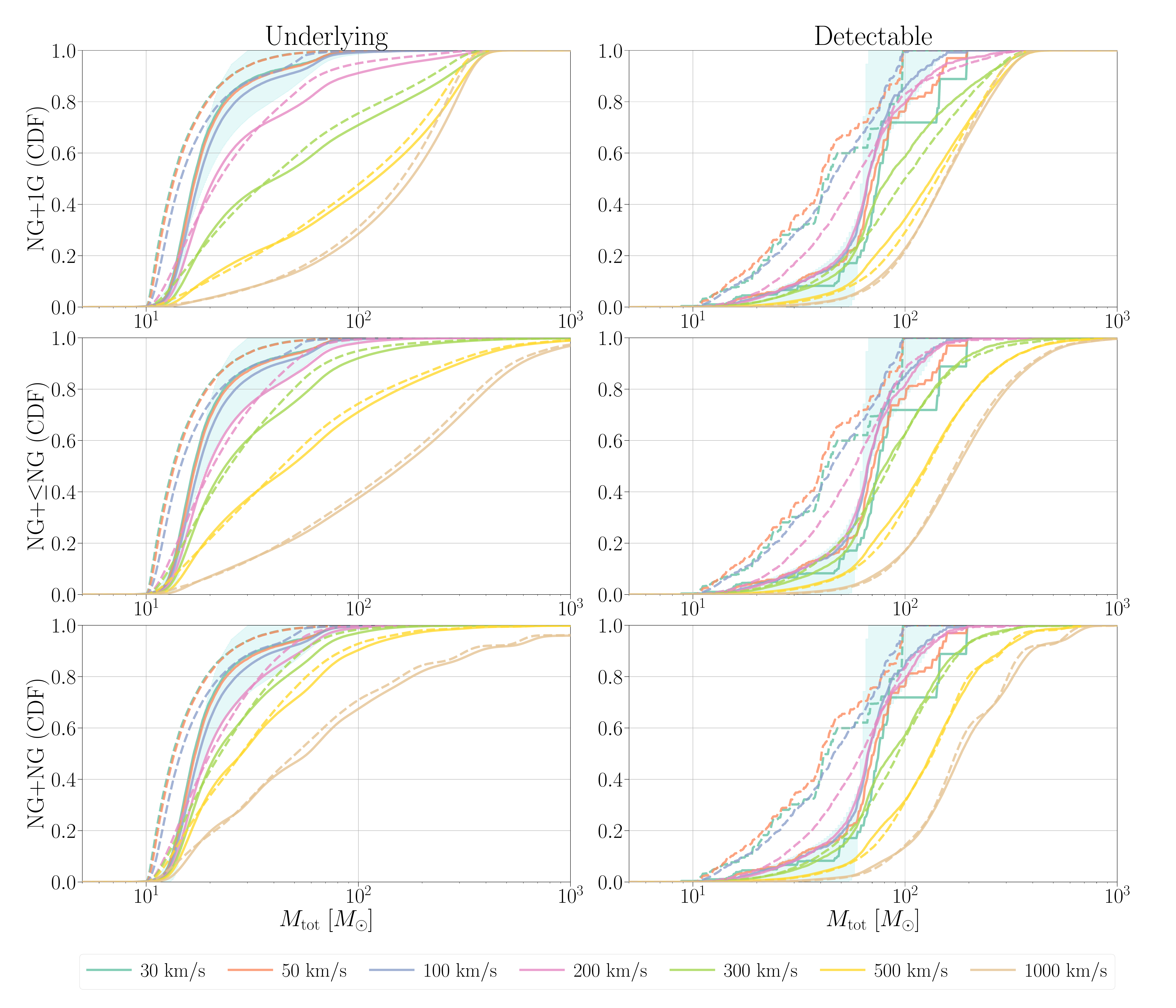}
\caption{Cumulative distributions for the underlying (left) and detectable (right) \ac{BBH} merger mass distributions using the three pairing methods of hierarchical mergers (top, middle, and bottom). 
Solid lines use the \LVC and dashed lines use the \IMF \OneG population model, with different colors denoting different escape velocities of the dynamical environment. 
Across pairing scenarios and \OneG population assumptions, hierarchical assembly in our models with escape velocities $\gtrsim 300~\kms$ predict more high-mass mergers than the constraints by the \ac{LVC}. 
}
\label{fig:escape_velocity_BHbudget_mass_cdf}
\end{figure*}

Contrary to the \NGNG pairing scenario, the retention fractions of \NGOneG and \NGMG scenarios tend to level out after a number of subsequent mergers have been lucky enough to be retained in their host environments. 
This behavior is particularly apparent for high escape velocities in the \NGOneG pairing because the buildup of the merger product mass will result in extreme mass ratios that receive negligible kicks upon merging with a member of the \OneG population. 
For the \NGOneG pairing and the \LVC (\IMF) \OneG population at escape velocities of $\gtrsim100~\kms$, $\fretMinOneThousandNGOneGLVC$ ($\fretMinOneThousandNGOneGIMF$) of merger products are retained forever, never having been ejected from their host environment. 
This percentage is still substantial for escape velocities down to $300~\kms$, with $\fretMinFiveHundredNGOneGLVC$ ($\fretMinFiveHundredNGOneGIMF$) and $\fretMinThreeHundredNGOneGLVC$ ($\fretMinThreeHundredNGOneGIMF$) of systems never being ejected for environments with escape velocities of $500~\kms$ and $300~\kms$, respectively. 

The coevolution of mass growth and recoil kick strength as the number of mergers increases leads to the efficient buildup of merger product mass for moderate to high escape velocities in each pairing scenario considered. 
For \NGOneG pairing, the rate of mass growth of the hierarchical merger product is slower, though the retention efficiency is much higher and can lead to long chains of subsequent mergers that can cause significant buildup in mass. 
Though less efficient at retaining merger products, the mass buildup in the \NGNG pairing is expedited due to pairings that tend to be near equal in mass. 
The \NGMG scenario exhibits a mix of both of these general features. 
In the following subsection, we turn to the expected mass distributions of hierarchically assembled \acp{BH} when escape velocities are incorporated and compare them with the empirical mass distribution from current \ac{GW} observations.

\subsection{Mass Distributions through Hierarchical Assembly}\label{subsec:results_mass_distributions}

We now examine the resultant mass distributions of our hierarchically assembled \acp{BH} using various assumptions for the escape velocity of their host environments. 
Mirroring the results of Figure~\ref{fig:retention_fraction}, the total mass distributions in Figure~\ref{fig:escape_velocity_BHbudget_mass_cdf} push to larger values as the escape velocities of their hosts increase due to longer chains of hierarchical mergers being retained. 
For example, using the \LVC \OneG population and the \NGOneG pairing, we find that \NGOneGGreaterEightyThreeHundred of hierarchical mergers have a source-frame total mass $> 80~\Msun$ for escape velocities of $300~\kms$, \NGOneGGreaterEightyFiveHundred for escape velocities of $500~\kms$, and \NGOneGGreaterEightyOneThousand for escape velocities of $1000~\kms$. 

Due to the Malmquist bias inherent to compact binary mergers, more massive mergers are more luminous, which pushes the detectable mass distribution to higher values. 
However, due to seismic noise that impinges Earth-based detectors, \acp{BBH} above a certain detector-frame mass will merge at too low of a \ac{GW} frequency to be detected. 
Thus, the \acp{CDF} of the detectable distributions plateau at $\Mtot \gtrsim 500~\Msun$. 
For the \LVC \OneG population and the \NGOneG pairing, we find that \NGOneGGreaterEightyThreeHundredDet, \NGOneGGreaterEightyFiveHundredDet, and \NGOneGGreaterEightyOneThousandDet of detectable hierarchical mergers have source-frame total masses $> 80~\Msun$ for escape velocities of $300~\kms$, $500~\kms$, and $1000~\kms$, respectively. 

Mass distributions push to lower values when considering the \IMF \OneG population as opposed to the \LVC \OneG population. 
This is due to the \IMF population having a steeper power-law slope and truncating at a lower maximum mass. 
Though natal spins are lower for the \IMF \OneG population (and thus kicks are suppressed, see the bottom row of Figure~\ref{fig:merger_product_evolution}), initial spins are mostly washed out after the first merger and the drop of retention efficiency for subsequent mergers follows a similar behavior as the \LVC model (Figure~\ref{fig:retention_fraction}).

% FIGURE 4
\begin{figure}[t]
\begin{center}
\includegraphics[width=0.42\textwidth]{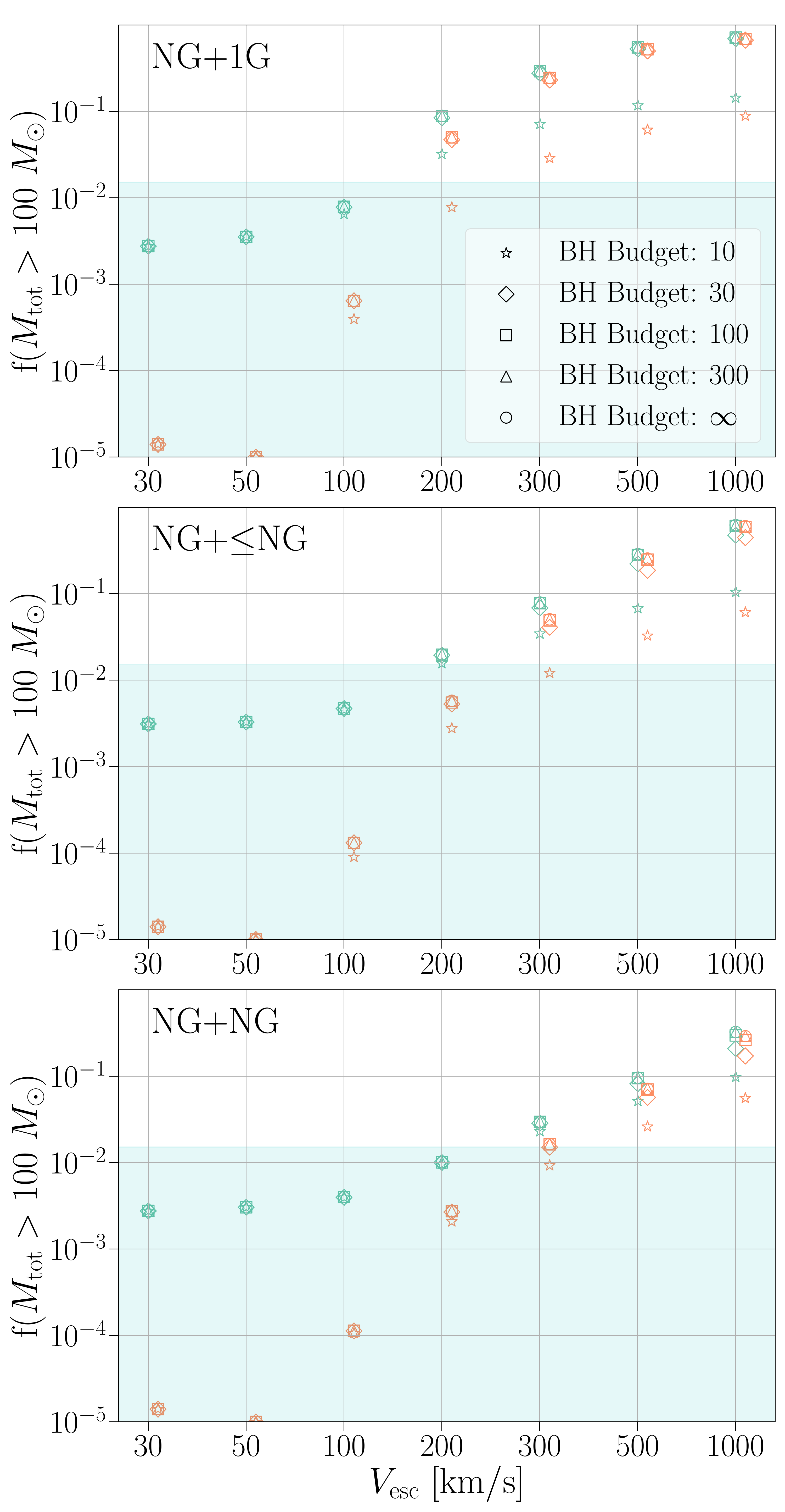}
\caption{Expected fraction of hierarchical \ac{BBH} mergers with a source-frame total mass greater than $100~M_\odot$ as a function of escape velocity for the three hierarchical merger pairing models. 
Green and orange markers denote the \LVC and \IMF \OneG pairings, respectively, with marker styles denoting different assumptions for the total \ac{BH} budget available for hierarchical assembly. 
The cyan shaded region marks the 90\% credible interval of the posterior predictive distribution from the \texttt{Power Law + Peak} model; the lower end of this interval stretches below the vertical bounds of this figure. 
Markers for the \IMF \OneG population are artificially offset for readability. 
Points that lie on the horizontal axis indicate that $f(M_\mathrm{tot} > 100 \Msun) < 10^{-5}$.
Points that lie above the cyan shaded region overproduce massive \acp{BH} if hierarchical mergers exclusively account for the full \ac{BBH} population; this is observed in most models where escape velocities are $\geq 300~\kms$. 
Reducing the \ac{BH} budget available for hierarchical assembly only has a noticeable effect at values of $\lesssim 30$. 
}
\label{fig:expected_fraction_above_mass}
\end{center}
\end{figure}

We emphasize that these mass distributions only consider \ac{BBH} mergers formed through hierarchical assembly, thus ignoring the potential additional contribution of \OneGOneG mergers in the same dynamical environment as well as other formation channels that may contribute to the full \ac{BBH} population, such as mergers that result from isolated field binaries. 
Therefore, the level of inconsistency between the empirical \ac{LVC} distribution and the mass distributions we construct act to jointly constrain the branching fraction of a particular dynamical formation channel as well as the fraction of \ac{BBH} mergers within that channel that are the result of hierarchical assembly relative to \OneGOneG mergers. 

Figure \ref{fig:expected_fraction_above_mass} shows the expected fraction of mergers above a source-frame total mass of $100~\Msun$ as a function of escape velocity predicted by our hierarchical merger models. 
The cyan band indicates the number of \ac{BBH} mergers with a total source-frame mass $> 100~\Msun$ based on the posterior predictive distribution measured using the \texttt{Power Law + Peak} mass model in the \ac{LVC} analysis~\citep{GWTC3_pops}; \LVCLessThanOneHundred of \ac{BBH} mergers have a total source-frame masses greater than $100~\Msun$ at 95\% credibility. 
For most pairing models, once escape velocities reach ${\sim 300~\kms}$, the fraction of hierarchically assembled systems with total masses greater than $100~\Msun$ exceeds the upper limit measured in the \LVC mass distribution. 
The exception is the \NGNG pairing model with the \IMF \OneG population, where even at an escape velocity of $300~\kms$ the percentage of hierarchically assembled systems with $M_\mathrm{tot} > 100 \Msun$ is \FracAboveOneHundredMsunNGNGIMFThreeHundred. 
Pairing methods that prefer symmetric masses in mergers always have a lower fraction of high-mass systems; though the buildup in mass is expedited (Figure~\ref{fig:merger_product_evolution}), recoil kicks are consistently higher and merger products are more readily ejected (Figure~\ref{fig:retention_fraction}) which suppresses the high end of the mass spectrum. 
For example, at an escape velocity of $500~\kms$ the percentage of hierarchically assembled mergers with source-frame total masses greater than $100~\Msun$ is \FracAboveOneHundredMsunNGNGLVCFiveHundred, \FracAboveOneHundredMsunNGMGLVCFiveHundred, and \FracAboveOneHundredMsunNGOneGLVCFiveHundred for \NGNG, \NGMG, and \NGOneG, respectively. 

Another parameter that acts to suppress the high end of the mass spectrum is the overall budget of \acp{BH} available for hierarchical assembly. 
The effect of imposing a \ac{BH} budget is shown with different symbols in Figure~\ref{fig:expected_fraction_above_mass}, while the impact of this parameter on the total mass distributions is shown in Figure~\ref{fig:escape_velocity_mass_cdf_BHbudget_LVC} of Appendix~\ref{app:BH_budget}. 
Other than the lowest assumed \ac{BH} budget of $10$, imposing a \ac{BH} budget has no effect on the \NGOneG pairings because larger assumed values still exceed the maximum number of \acp{BH} needed to fully grow an \NGOneG branch. 
For the \NGMG and \NGNG pairings, the impact of imposing a \ac{BH} budget is most apparent at higher escape velocities. 
The assumed \ac{BH} budget has the largest effect on the \NGMG pairing because recoil kicks are suppressed relative to the \NGNG pairing, and thus, this constraint acts to suppress mass growth at the highest escape velocities. 
At an escape velocity of $1000~\kms$, the percentage of systems in the \NGMG pairing scenario and the \LVC \OneG population with total masses greater than $100~\Msun$ drops from \FracAboveOneHundredMsunNGMGLVCOneThousandBHBudgetOneHundred for a \ac{BH} budget of $100$ to \FracAboveOneHundredMsunNGMGLVCOneThousandBHBudgetThirty for a \ac{BH} budget of $30$ and \FracAboveOneHundredMsunNGMGLVCOneThousandBHBudgetTen for a \ac{BH} budget of $10$.

\section{Discussion}\label{sec:discussion}

Hierarchical \ac{BBH} mergers are a natural occurrence in dynamical environments, with merger rates strongly dependent on the escape velocity of their host. 
We demonstrate that a range of models inevitably overproduce high-mass mergers through hierarchical assembly, leading to tension with current observational constraints if such formation environments dominate the \ac{BBH} merger rate. 
Using simple models to represent \ac{BBH} pairing functions and properties of host environments, we explore the retention efficiency and resultant mass spectrum of \acp{BBH} in the hierarchical merger paradigm. 
Regardless of whether host environments preferentially lead to intergenerational (\NGOneG and \NGMG) or equal-generation (\NGNG) pairings, we find hierarchical mergers in environments with moderate to high escape velocities (${\sim 300\mbox{--}1000~\kms}$) will efficiently produce high-mass \ac{BBH} mergers, with a significant fraction having total source-frame masses $\gtrsim 100~\Msun$. 
For environments where intergenerational pairings are preferred, the mass buildup is slower, though the retention fraction eventually flattens out and leads to a runaway merger scenario. 
For equal-generation pairings, the retention fraction continually drops with each merger generation, but the mass buildup is significantly expedited. 
Assuming a large budget of \acp{BH} available for hierarchical mergers, environments with escape velocities $>500~\kms$ result in at least $\MassesAboveOneHundredMsunFiveHundredRange$ of their hierarchical mergers having source-frame total masses greater than $100~\Msun$. 

Depending on the relative number of hierarchical mergers compared to \OneGOneG mergers, such environments therefore exhibit potential incompatibility with the observed \ac{BBH} mass spectrum due to an overabundance of high-mass mergers, particularly if such environments contribute significantly to the local \ac{BBH} merger rate. 
\acp{NC} with escape velocities of $\gtrsim300~\kms$ are rare in the local universe~\citep{Antonini2016a} and thus likely have a minor contribution to the observed population of \acp{BBH}. 
Local \ac{BBH} merger rate predictions from \ac{AGN} disks span many orders of magnitude due to the many physical uncertainties inherent to this channel~\citep{McKernan2018,Grobner2020}, with certain studies arguing that mergers from \ac{AGN} disks make up a significant fraction ($>25\%$) of the detected \ac{BBH} population~\citep[e.g.,][]{Ford2021,Gayathri2021}. 
We note that multichannel analyses of dynamical environments such as \cite{Mapelli2022}, which consider cosmologically motivated rate estimates of both \OneGOneG and hierarchical mergers from young star clusters, \acp{GC}, and \acp{NC} in addition to field binaries, find no overproduction of hierarchical mergers, though the \ac{AGN} disk scenario is not included in their analysis. 

We stress that our analysis does not directly constrain branching fractions from dynamical environments; rather, it provides an estimate of the joint constraints on branching fractions and the relative ratio of \OneGOneG mergers to hierarchical mergers. 
For example, in environments with escape velocities of ${\sim 1000~\kms}$, an \NGOneG pairing function, and an \IMF \OneG population, similar to what one might expect for an \ac{AGN} disk scenario~\citep[see, e.g.,][]{Yang2019b,Tagawa2021}, we find \FracAboveOneHundredMsunNGOneGOneThousandIMF of hierarchical mergers to have total masses greater than $100~\Msun$, whereas the population inference results from the \ac{LVC} find that \LVCLessThanOneHundred of all mergers have masses greater than $100~\Msun$ at $90\%$ credibility~\citep{GWTC3_pops}. 
If one were to assume mergers in \ac{AGN} disks dominate the production of systems observed by the \ac{LVC}, this provides a rough estimate that \OneGOneG mergers must be at least \AGNFirstGenerationToHierarchical times more prevalent than hierarchical mergers. 
Constraints can thus be placed on aspects of \ac{AGN} environments that influence the relative rate of hierarchical mergers, such as disk lifetimes~\citep{Secunda2019} and how efficiently the orbits of hierarchical merger products decay into the central supermassive \ac{BH}. 
Alternatively, the high efficiency of forming high-mass systems through hierarchical mergers in environments with escape velocities $\gtrsim 300~\kms$ may indicate that these environments are not dominant contributors to the overall population of merging \acp{BBH}. 

The toy models considered in this paper are approximations of the detailed evolution of \ac{BH} mergers in dynamically rich environments. 
They may fall short of capturing important intricacies of hierarchical assembly, especially in environments that are gas rich or involve interactions with a central massive \ac{BH}. 
However, the multiple pairing methods we construct broadly capture the diversity of expected hierarchical assembly processes in various host environments. 
Additionally, we consider a single initial \ac{BH} mass function for all environments, whereas environments with low escape velocities may lead to a more top-heavy mass spectrum because low-mass \acp{BH} may be more readily ejected at formation by supernova kicks~\citep{Mapelli2021}. 
An analysis that self-consistently models the parameter distributions of both \OneG \acp{BH} and an arbitrary number of higher-generational populations could hold promise~\citep[e.g.,][]{Doctor2020,Kimball2020,Kimball2021,Mould2022}, though expanding such methodologies beyond second-generation hierarchical mergers is nontrivial and is the aim of future work. 
Nevertheless, studies such as \cite{Mould2022} have made headway in this direction by relying on machine-learning techniques to interpolate between astrophysical models, which they use to extend inference capabilities to account for higher-generational hierarchical mergers. 

We note that the formation of \acp{IMBH} through hierarchical assembly also requires that the growing \ac{BH} seeds pass through the sensitive frequency range of ground-based \ac{GW} detectors~\citep[e.g.,][]{Fragione2022c}. 
The detection, or nondetection, of mass-gap \acp{BH}, as well as \acp{BH} beyond the \ac{PI} mass gap, provide important constraints to \ac{IMBH} growth mechanisms. 
Supermassive \ac{BH} growth may also be due in part to the hierarchical assembly of stellar-mass \ac{BH} seeds~\citep[e.g.,][]{Chen2022}, and in this scenario, one would also expect to observe such mergers at high redshifts in the sensitive frequency range of future ground-based \ac{GW} detectors. 

Our work highlights the tension between models that produce binaries in and above the \ac{PI} mass gap at high efficiency and existing limits on mergers beyond the mass gap. 
The high efficiencies by which a particular formation channel produces extreme systems may be detrimental to its consistency with the population of \ac{GW} sources as a whole. 
Careful consideration of overall population properties and selection effects, as well as the inclusion of multiple potential formation channels, are necessary to contextualize extreme systems with respect to the full population of \ac{GW} signals. 
Viable models for the formation of stellar-mass \acp{BH}, \acp{IMBH}, and supermassive \ac{BH} seeds may need to populate the \ac{PI} mass gap while carefully avoiding overpopulating the mass region immediately above the gap. 
Without care, models that populate the gap can lead to a cluster catastrophe, producing runaway growth of \acp{BH} with masses $\gtrsim 100~M_\odot$, in tension with existing observational limits.

\acknowledgments
We thank Maya Fishbach and Shanika Galaudage for providing code for drawing mock events from population models. 
We acknowledge Will Farr, Saavik Ford, Giacomo Fragione, Chase Kimball, Kyle Kremer, Michela Mapelli, Barry McKernan, and Carl Rodriguez for useful discussions. 
We also thank the anonymous referee whose comments significantly improved this manuscript. 
Support for this work and M.Z. was provided by NASA through the NASA Hubble Fellowship grant HST-HF2-51474.001-A awarded by the Space Telescope Science Institute, which is operated by the Association of Universities for Research in Astronomy, Incorporated, under NASA contract NAS5-26555. 
D.E.H. is supported by NSF grants PHY-2006645, PHY-2011997, and PHY-2110507, as well as by the Kavli Institute for Cosmological Physics through an endowment from the Kavli Foundation and its founder Fred Kavli. 
This work was performed in part at the Aspen Center for Physics, which is supported by NSF grant PHY-1607611.
This material is based upon work supported by NSF's LIGO Laboratory which is a major facility fully funded by the National Science Foundation.

\software{\texttt{Astropy}~\citep{TheAstropyCollaboration2013,TheAstropyCollaboration2018}; 
\texttt{Bilby}~\citep{Ashton2019}; 
\texttt{iPython}~\citep{ipython}; 
\texttt{Matplotlib}~\citep{matplotlib}; 
\texttt{NumPy}~\citep{numpy,numpy2}; 
\texttt{Pandas}~\citep{pandas}; 
\texttt{precession}~\citep{Gerosa2016}; 
\texttt{SciPy}~\citep{scipy}.}

\clearpage
\appendix

\section{Variations to Delay Times in Hierarchical Assembly}\label{app:delay_times}

%%% FIGURE 5 %%%
\begin{figure*}[b]
\includegraphics[width=1.0\textwidth]{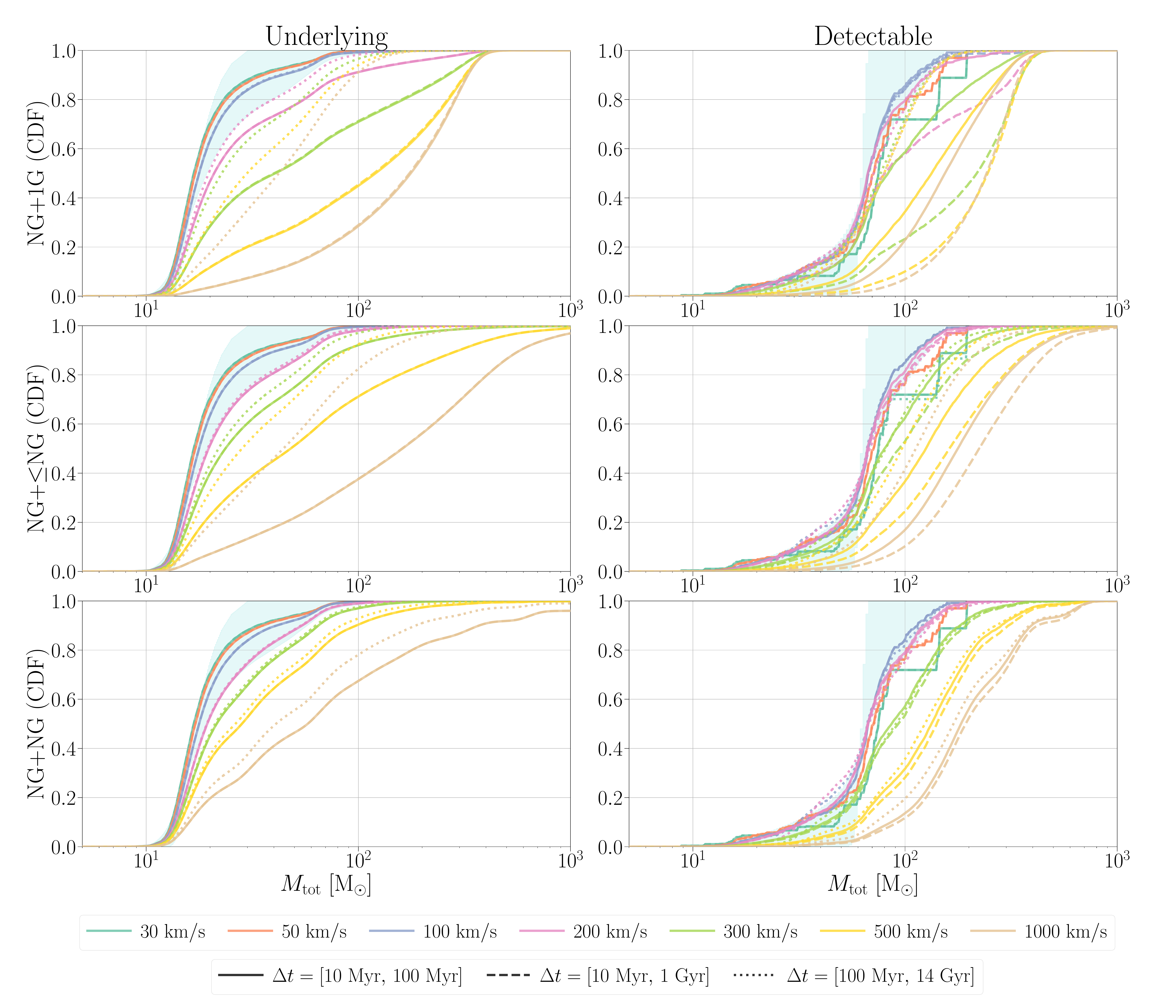}
\caption{Same as Figure~\ref{fig:escape_velocity_BHbudget_mass_cdf}, except with differing line styles showing the mass distribution for various assumptions regarding the model for delay times between subsequent mergers. 
Slightly extending the delay-time distribution to $\Delta t = [10~\mathrm{Myr},\ 1~\mathrm{Gyr}]$ has virtually no effect on the underlying mass distribution and pushes the detectable distribution to slightly higher masses due to more systems merging within the horizon of \ac{GW} detectors. 
The long delay-time distribution of $\Delta t = [100~\mathrm{Myr},\ 14~\mathrm{Gyr}]$ pushes both the underlying and detectable hierarchical merger mass spectra to lower masses due to higher-generational merger products preferentially merging beyond the present day. 
Only the \LVC \OneG population model is shown for clarity. 
}
\label{fig:escape_velocity_mass_cdf_Tdel_LVC}
\end{figure*}

In Section~\ref{subsec:merger_trees} we describe the simple model used for initializing a particular merger tree at a given redshift, as well as our assumptions for the delay time between subsequent mergers. 
The fiducial model we assume for delay times, which is a log-uniform distribution between $10~\mathrm{Myr}$ and $100~\mathrm{Myr}$, is representative of gas-free environments with relatively short mass segregation timescales. 
Here, we explore variations in the assumed delay-time distribution, extending our fiducial distribution to log-uniform between $\Delta t = [10~\mathrm{Myr},\ 1~\mathrm{Gyr}]$, as well as considering a log-uniform between $\Delta t = [100~\mathrm{Myr},\ 14~\mathrm{Gyr}]$. 
Figure~\ref{fig:escape_velocity_mass_cdf_Tdel_LVC} shows how these variations affect the underlying and detectable \ac{BH} mass spectrum of hierarchical mergers across our range of assumed escape velocities.

%%% FIGURE 6 %%%
\begin{figure*}[b]
\includegraphics[width=1.0\textwidth]{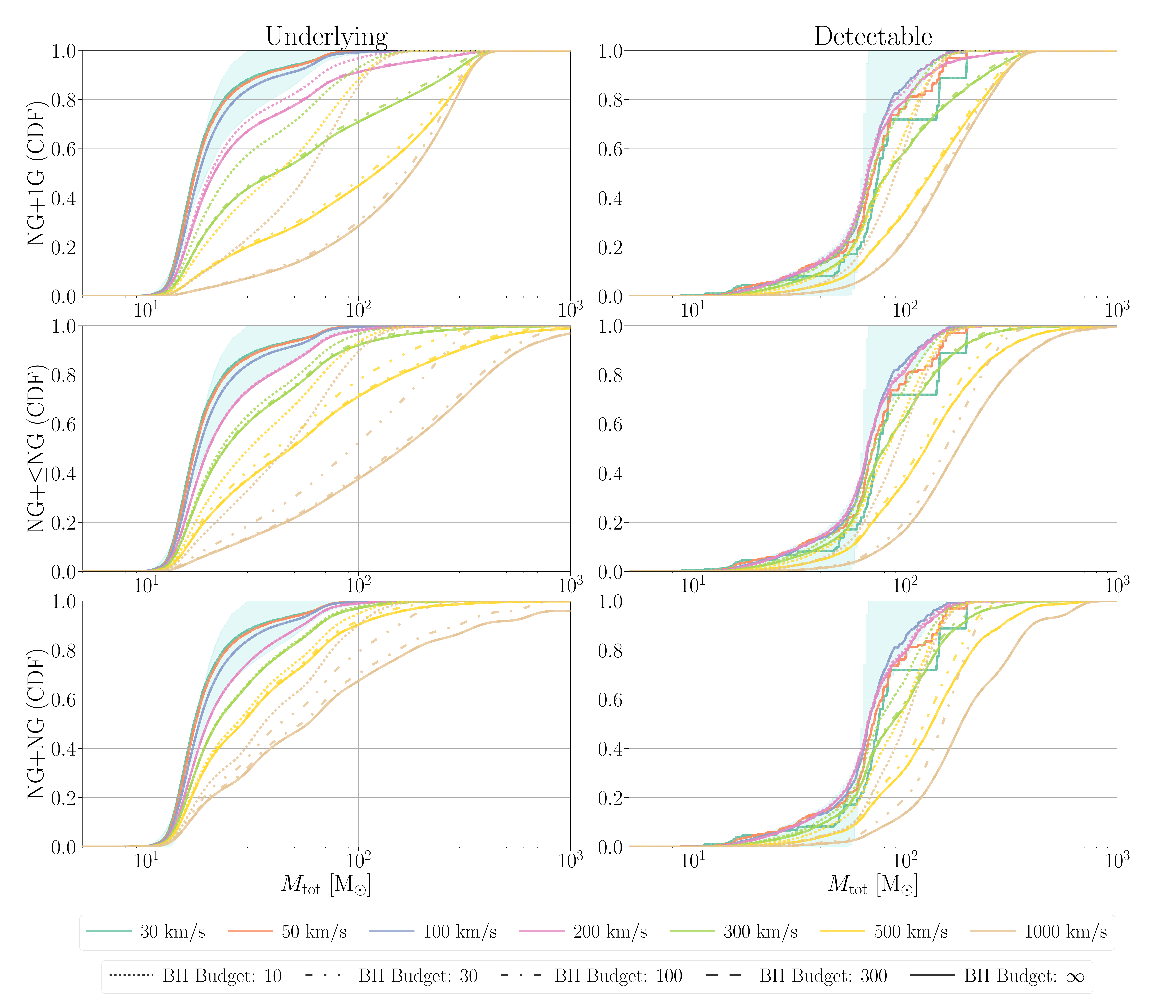}
\caption{Same as Figure~\ref{fig:escape_velocity_BHbudget_mass_cdf}, except with differing line styles showing the mass distribution for various assumptions regarding the total budget of \acp{BH} available for mergers. 
Lowering the total \ac{BH} budget available for hierarchical assembly naturally pushes the hierarchical merger mass spectra to lower masses as it truncates the hierarchical assembly after less repeated mergers, though this parameter has little impact on the mass spectrum so long as the \ac{BH} budget is $\gtrsim 30\mbox{--}100$, depending on the pairing scenario. 
Only the \LVC \OneG population model is shown for clarity. 
}
\label{fig:escape_velocity_mass_cdf_BHbudget_LVC}
\end{figure*}

Increasing the assumed delay times to $\Delta t = [10~\mathrm{Myr},\ 1~\mathrm{Gyr}]$ has a negligible impact on the underlying distribution, as the increase in the number of systems that now merge at times later than the present is minor. 
This impact is more prominent with the $\Delta t = [100~\mathrm{Myr},\ 14~\mathrm{Gyr}]$ delay-time model because many more systems now readily merge after the present day (dotted lines in Figure~\ref{fig:escape_velocity_mass_cdf_Tdel_LVC}). 
This acts to truncate the mass distribution, as the most massive mergers that have proceeded through the most hierarchical mergers more readily merge at times beyond the present. 
The detectable distributions in the right-hand column of Figure~\ref{fig:escape_velocity_mass_cdf_Tdel_LVC} show more complex behavior. 
For all pairing scenarios and escape velocities, the extended $\Delta t = [10~\mathrm{Myr},\ 1~\mathrm{Gyr}]$ delay-time model pushes to larger source-frame total masses in the detectable distribution than the fiducial model. 
This is due to high-mass systems from numerous chains of hierarchical mergers more readily merging in the local universe and being more accessible to ground-based \ac{GW} observatories. 
On the other hand, the longest delay-time model explored ($\Delta t = [100~\mathrm{Myr},\ 14~\mathrm{Gyr}]$) truncates the high end of the mass spectrum significantly due to many more systems merging beyond the present day. 
This is especially apparent in the \NGOneG pairing model, as a large quantity of subsequent hierarchical mergers is necessary to significantly build up the mass of the merger product. 
The influence of large delay times is less apparent in the detectable distributions for near-equal-mass pairings, such as the \NGNG pairing scenario, as it requires a smaller number of repeated mergers to bolster the mass spectrum significantly.

\section{Growing on a Budget}\label{app:BH_budget}

For mass distributions presented in the main text, we assume that the number of \acp{BH} the hierarchical merger product can consume is only limited by the maximum number of mergers possible for each pairing scenario as described in Section~\ref{subsec:merger_trees}. 
Here, we relax this assumption by imposing a limit on the \ac{BH} budget available for each hierarchical merger branch. 
This accounts for all \acp{BH} consumed throughout the evolution of the merger product. 
Figure \ref{fig:escape_velocity_mass_cdf_BHbudget_LVC} displays variations in the resultant mass distribution of hierarchical mergers for the \LVC \OneG population model when differing assumptions for the \ac{BH} budget are imposed. 
We note that a more physically realistic approach for implementing the \ac{BH} budget would depend on the escape velocity of the host environment, which in turn depends on the total mass of the cluster environment. 
However, in this analysis, we keep these parameters uncorrelated to explore their impact on mass distributions independently.

\bibliography{library}{}
\bibliographystyle{aasjournal}

\end{document}